\documentclass[aps,prc,tightenlines,groupedaddress,nofootinbib,
showpacs,preprintnumbers,amsmath,amssymb,superscriptaddress]{revtex4}
\usepackage{graphicx}     
\usepackage{dcolumn}     
\usepackage{bm}              
\usepackage{amsmath}
\begin{document}

\title{Convergence of 2-Body Effective Range Expansions for $nd$ Quartet Scattering}
\author{J. R. Shepard}
\affiliation{Department of Physics, University of Colorado, 
	Boulder CO 80309}
\author{J. A. McNeil}
\affiliation{Department of Physics, Colorado School of Mines, 
	Golden CO 80401}
\date{\today} 
\begin{abstract}
We examine the convergence properties of the 2-nucleon Effective Range Expansion as used in Effective Theories (ET-ERE's) for 3-nucleon calculations. We accomplish this by accounting for the 2-body dynamics with a simple rank-1 separable 2-body potential where the finite range effects can be incorporated systematically in both the 2- and 3-body problems.  We make our initial comparisons in the simple context of the  $^3S_1$ 2-nucleon channel and the $^4S_{3/2}$  3-nucleon channel.  We find that convergence problems for some of the 3-nucleon scattering amplitudes using the ET-ERE can be traced to its poor account of finite range effects that soften the momentum dependence of the deuteron propagator in the Faddeev kernel.  In contrast, our simple separable potential with dipole form factors works very well in all cases considered. 

\end{abstract}
\pacs{21.45.-v,25.40.Dn,34.50.-s}
\maketitle 

\section{Introduction}
\label{Sec:Intro}
Much attention has recently been paid to the low-energy physics of the neutron-deuteron ($nd$) system using the powerful methods of Effective Theories (ET's)~\cite{Lepage:1990,Georgi:1994,Kaplan:1995uv,Kaplan:1996xu,Kaplan:1998we,vanKolck:1999mw}. The basis of these approaches is the assumption that, if few body systems display long distance features, these should not depend on details of the short distance physics. More specifically, most treatments are based on a field theory specified by a Lagrangian which contains a series of contact interactions with coefficients 
depending on increasing powers of momentum. This series, it is hoped, may be truncated at some low order and still account for the low energy physics once the (few) parameters of the ET  are fixed by external inputs. The power of the approach is that these few inputs yield a theory which can then be used to calculate systematically any low energy property of the system.

ET's are very powerful in the limit that the 2-body scattering length is much larger than any other length scale in the problem. It is clear that, in this case, microscopic details of the 2-body physics are irrelevant and that any solutions found are ``universal'' in nature; {\it i.e.}, are applicable to any system for which the limit is valid. In such cases, universality arises in the ET's because only the lowest order contact interaction is needed. It is naturally of interest to extend this powerful formulation to problems for which the  scattering length is large, but not overwhelmingly so compared to other length scales characterizing the 2-body interaction. A common means of identifying the appropriate 2-body length scales is the Effective Range Expansion (ERE)~\cite{Bethe:1949}, namely
\begin{equation}
p \cot\delta=-\frac{1}{a}+\frac{r_0}{2} p^2+v_4 p^4\dots,
\label{ERE0}
\end{equation}
where $p$ is the 2-body relative momentum, $\delta$ is the scattering phase shift at momentum $p$, $a$ is the scattering length, $r_0$ is the effective range, $v_4$ is the shape parameter, and higher order terms are implicit. When applying ET's to low energy phenomena, it is basically the effective range which governs modifications to universal results as one leaves the universal regime. Such modifications may be referred to generically as ``finite range effects'', whether or not they can be wholly accounted for by $r_0$ alone. In the present work, we examine finite range effects in the low energy 3-body problem, focussing on quartet $nd$ scattering, to which ET's have been applied repeatedly over the last decade~\cite{Bedaque:1997qi, Bedaque:1998mb, Bedaque:1998km, Griesshammer:2004pe, Platter:2006ev}.  Several of these previous works have addressed finite range effects in a manner tied closely to the ERE as discussed above. 

Here we adopt a different approach which incorporates the finite range of the 2-body interactions at the outset. More specifically, we treat  the 2-body interaction as a sum of separable terms~\cite{Harms:1970}, each consisting of the outer product of  {\it form factors} which reflect the structure of interactions which are directly related to real or virtual 2-body states. The use of separable potentials leads to considerable simplification. Specifically, separable potentials admit an algebraic solution to the two-body problem greatly simplifying the three-body integral equations.  For the specific purposes of this study, the effective range expansion is seen as the zero-range limit of the separable expression whose expansion in the separable potential's range parameter illuminates the role of finite range effects allowing a critical examination of the manner in which effective range expansions account for finite range effects in the three-body context.

Separable potentials have a long history in nuclear physics going back to the period immediately following Faddeev's seminal work on the 3-body problem (see Refs.~\cite{Watson:1967} and~\cite{Lovelace:1964} for comprehensive lists of representative papers) when calculations of $Nd$ scattering relied on finite-range separable NN interactions to an extent that was as nearly universal as the use of contact interactions in ET's is today (see, for example, Refs.~\cite{Lovelace:1964, Aaron:1964, Alt:1970}).  Use of this approach continues today, though on a much more limited basis. Examples in nuclear physics include References \cite{Koike:1986,Koike:1990,Kwong:1997,Kamada:2005hr}. The method has also been applied quite recently to the new problem of cold, dilute atomic gases in, {\it e.g.}, References \cite{Penkov:2003,Kohler:2003,Stoll:2005,Kohler:2006zz,shepard:062713}.

In the present work we exploit separable 2-body interactions to illuminate modern treatments of 3-body systems using effective theories which begin by assuming contact interactions and then rely on the Effective Range Expansion to account for the 2-body dynamics.  Several years ago, Lepage \cite{Lepage:1997cs} provided a general approach to developing effective interactions starting with the now-canonical series of contact interactions. However, in his treatment he let these morph into a similar series smeared in configuration space by what amounts to a form factor.  The possibility of using separable potentials is also mentioned in passing in several of the important papers in the recent history of ET's addressing three-body physics including one by Afnan and Phillips \cite{Afnan:2003bs}.  In any case, there is much current interest in how best to treat finite range effects that may enter in Efimov physics which is now accessible experimentally in cold atom systems where one exploits Feshbach resonances to tune the 2-body scattering length, $a$, to essentially any desired value. As the scattering length diverges, finite range effects become negligible and the physics is universal~\cite{Braaten:2004rn,Braaten:2006vd}. However, when away from this limit and when interested in how such a limit is approached with $a$ finite, it would seem prudent to use an approach that does not implicitly assume range effects are absent at leading order.

Building on earlier work by Weinberg~\cite{Weinberg:1963a}, E. Harms \cite{Harms:1970} developed a separable expansion for the 2-body potential which admits an algebraic matrix solution for the $T$-matrix.  Several authors have studied the convergence of the separable expansion of the 2-body potential in 3-body applications~\cite{Harms:1970a,Osman:1979,Haidenbauer:1986hk,Koike:1990}. The separable expansion truncated at leading order is historically referred to as the ``Unitary Pole Approximation" (UPA). In the present work we focus on the $^3S_1$  NN channel and implement the UPA using a dipole form factor with parameters that give a pole in the 2-body $T$-matrix at the 2-body binding energy (of the deuteron) and show how its associated form factor leads directly to the 2-body bound state wave function.  To establish this methodology in a simple context, we treat the 2-body channel without tensor coupling 
and then  focus on the $nd$ $^4S_{3/2}$ (quartet) 3-body channel
which allows us to avoid complications due to Efimov physics.  (Extensions to tensor coupling and the $nd$ doublet case will be treated elsewhere.)  We then develop a systematic expansion of the 3-body scattering amplitude in inverse powers of the dipole range parameter and demonstrate how the ET-ERE based on contact interactions follows as a limiting case of our model which further allows us to study its convergence properties in the 3-body context.  This exercise demonstrates some significant limitations of the ERE in accounting for finite range effects.

The paper is organized as follows. In Section II we review Harm's separable expansion. We focus on the rank-1 approximation, the UPA, for the NN problem in the $^3$S$_1$ channel and constrain its single inverse range parameter, $\beta$, using the Nijmegen II \cite{Nijmegen:NNOnline} phase shifts.  In Section III we use the UPA to fix the 2-body input to the 3-body Faddeev equations for $nd$ scattering in the $^4$S$_{3/2}$ and $^4$P$_J$ channels.  To provide a basis for examining finite range effects, we develop an expansion of the resulting Faddeev equations essentially in powers of $\gamma/\beta$ where $\gamma$ is related to the 2-body binding energy, $B$, via $\gamma=\sqrt{mB}$.  In Section IV we outline the treatment of the same problem using ET methods. We review various common approaches for including effects of the finite range of the 2-body interactions based on the Effective Range Expansion (ERE) of $p\cot\delta$ for 2-body elastic scattering.  We refer to such calculations as ET-ERE results. Section V presents comparisons of ET-ERE results at Leading Order (LO), Next-to-Leading Order (NLO) and Next-to-Next-to-leading Order (NNLO) with the earlier UPA results similarly expanded in powers of the range of the form factor. We find, not surprisingly, that the two approaches are identical at lowest order.  Some differences appear at NLO but there are qualitative differences at NNLO. Moreover, for $s$- and $p$-wave quartet $nd$ scattering, we find that both expansions of the 3-body equations suffer some catastrophic failures at NLO and NNLO while the full UPA provides a excellent description of the phase shifts in this channel in spite of its simplicity.   We present our conclusions and future directions in the last section.

\section{Separable Dipole 2-Body Interaction}
\label{Sec:ERE}
As noted in the introduction, separable potentials have a long history in nuclear physics going back to the period immediately following Faddeev's breakthrough work on the 3-body problem when, for computational convenience, calculations of $nd$ scattering relied almost exclusively on separable NN interactions.  The essential feature of these approaches that we will exploit here is the explicit inclusion of the range of the 2-body interaction and its close association with the bound state wavefunction. Following Harms~\cite{Harms:1970}, consider the 2-body Schroedinger equation ($\hbar$, $c=1$, $m_{neutron}=m_{proton}=m$):
\begin{equation}
G_0^{-1}(E_n)|\chi_n\rangle =V|\chi_n\rangle,
\label{twobdySE}
\end{equation}
where $G_0(E)$ is the free 2-body propagator. Assume there exists a single bound state at $E\rightarrow -B$. The bound state satisfies
\begin{equation}
G_0^{-1}(-B)|\chi_B\rangle =V|\chi_B\rangle .
\label{twobdybnd1SE}
\end{equation}
Defining the {\it form factors} via the operation of $G_0^{-1}(-B)$ as
\begin{equation}
|\psi_n\rangle=G_0^{-1}(-B)|\chi_n\rangle,
\label{defFF}
\end{equation}
we see that the bound state Schroedinger equation can be written as
\begin{equation}
V G_0(-B)|\psi_B\rangle =\lambda_B |\psi_B\rangle ,
\label{twobdybnd2SE}
\end{equation}
where $\lambda_B=1$. Solutions $|\psi_n\rangle$ of the generalization of Eq.~\ref{twobdybnd2SE},
\begin{equation}
V G_0(-B)|\psi_n\rangle =\lambda_n |\psi_n\rangle,
\label{twoFFeqn}
\end{equation}
form a complete set of kets with normalization condition
\begin{equation}
\langle\psi_n|G_0(-B)|\psi_n\rangle=-\delta_{n,m} .
\label{FFnorm1}
\end{equation}
We call the $|\psi_n\rangle$'s the ``form factors'' of the potential $V$. (Note that our $\lambda_n$'s are the {\it inverses} of those defined by Harms.) 
 Also note that, with this normalization convention, the bound state wave function is {\it not} normalized. We may define the wavefunction renormalization $Z$ via
\begin{equation}
Z_B^{-1}=\langle\chi_B |\chi_B\rangle= \langle\psi_B|G_0(-B)^2|\psi_B\rangle=-\frac{d}{dE} \bigg|_{E=-B} \langle\psi_B|G_0(E)|\psi_B\rangle .
\label{Zbnd}
\end{equation}
When there is a single bound state, the eigenvalues satisfy
\begin{equation}
\lambda_{n\neq B}<\lambda_B =1.
\label{FFnorm2}
\end{equation}
We observe that the potential can be written as
\begin{equation}
V=-\sum_n\ |\psi_n\rangle \lambda_n \langle\psi_n|,
\label{vFFexp}
\end{equation}
which is verified by substitution into Eq.~\ref{twoFFeqn} and then using the normalization condition Eq.~\ref{FFnorm1}.
After ordering the terms in this summation by their magnitudes of eigenvalues beginning with $\lambda_1=1$,
 a ``rank-$N$" approximation to the 2-body potential results from truncating the sum at $N$ terms. The 2-body $T$-matrix satisfies the Lippmann-Schwinger (LS) equation,
\begin{equation}
T({\bf p},{\bf p}';E^+)=V({\bf p},{\bf p}')+\int\frac{d^3q}{(2\pi)^3}\ \frac{V({\bf p},{\bf q})\ T({\bf q},{\bf p}';E)}{E^+ -q^2/m}.
\label{LSeq1}
\end{equation}
It is easy to show that
\begin{equation}
T({\bf p},{\bf p}';E^+)=-\sum_{n,m=1}^N\  \langle{\bf p} |\psi_n\rangle\ \Delta(E^+)_{n,m}\ \langle\psi_m|{\bf p}'\rangle ,
\label{LSeq2}
\end{equation}
where
\begin{equation}
[\Delta^{-1}(E^+)]_{n,m}=\lambda_n^{-1} \delta_{n,m}+\langle\psi_n|G_0(E^+)|\psi_m\rangle .
\label{DeltaDef}
\end{equation}
Thus, we see that the 2-body T-matrix associated with a separable potential to a given rank involves inverting a matrix with dimension of that rank.  The Unitary Pole Approximation (UPA) corresponds to truncating the expansion of $V$ and $T$ at rank-one, including only the form factor for the bound state:
\begin{equation}
V\rightarrow -|\psi_B\rangle\  \langle\psi_B|,
\label{vUPA}
\end{equation}
and
\begin{equation}
T({\bf p},{\bf p}';E^+)\rightarrow -\langle{\bf p}|\psi_B\rangle\ \Delta(E^+)\ \langle\psi_B|{\bf p}'\rangle,
\label{LSeq3}
\end{equation}
where we have used $\lambda_B=1$ and where 
\begin{equation}
\Delta^{-1}(E^+)=1+ \langle\psi_B | G_0(E^+)|\psi_B\rangle.
\label{DeltaE}
\end{equation}
To proceed, we make a specific choice for the functional form of 
 the bound state form factor. We select a dipole (or Yamaguchi\cite{Yamaguchi:1954}) form, namely
\begin{equation}
\langle {\bf p}|\psi_B\rangle={\cal N}^{1/2} g(p^2) =\frac{{\cal N}^{1/2}}{(1+p^2/\beta^2)},
\label{DipoleFF}
\end{equation}
where $g$ is the form factor which satisfies $g(0)=1$, $p^2={\bf p}^2$, and $\cal N$ is found by normalizing according to Eq.~\ref{FFnorm1},
\begin{equation}
{\cal N}^{-1}=\int \frac{d^3p}{(2\pi)^3} \frac{g(p^2)^2}{B+p^2/m}=\frac{m}{8\pi}\frac{\beta}{(1+\gamma/\beta)^2},
\label{DipoleNorm}
\end{equation}
where $\gamma=\sqrt{m B}$ has been used.   Similarly, 
\begin{equation}
\Delta^{-1}(E^+)= 1-\biggl(\frac{\beta+\gamma}{\beta+\sqrt{-m E^+}}\biggr)^2.
\label{DeltaUPA}
\end{equation}
 Finally, using the dipole form factor in Eq.~\ref{Zbnd} gives the wave function normalization factor,
\begin{equation}
Z_B^{-1}=\frac{m}{\gamma(\beta+\gamma)} .
\label{ZUPA}
\end{equation}

It is clear that, because of the normalization condition, the 2-body $T$-matrix will have a pole at $E=-B$ as required. It is interesting to note that this choice of form factor means we are, in effect, using the Hulthen parameterization of the deuteron wave function\cite{Hulthen:1957}. This can be confirmed by using the inverse of Eq.~\ref{defFF}, namely,
\begin{equation}
\langle p|\chi_B\rangle=\langle p|G_0(-B)|\psi_B\rangle,
\end{equation}
to find the momentum space bound state wavefunction, $\langle p|\chi_n\rangle$. Fourier transforming to configuration space yields the radial wave function,
\begin{equation}
u(r)=r\ \chi_B(r)\propto ({\rm e}^{-\gamma r}-{\rm e}^{-\beta r}),
\end{equation}
which in turn implies a local $r$-space NN potential,
\begin{equation}
V(r)=-\frac{\beta^2-\gamma^2}{m\bigl[{\rm e}^{(\beta-\gamma)r}-1\bigr]}.
\end{equation}
We are reminded that, in the lore of the day when the Hulthen wave function was in widespread use, the rule of thumb was that $\beta\sim 6\gamma$ produced a good account of the structure of the deuteron.

Using Eqs.~\ref{LSeq3}, ~\ref{DeltaE}, and ~\ref{DipoleFF}, we may construct a fully off-shell elastic 2-body amplitude, $f_2$, via
\begin{equation}
f_2(p,p';E)=-\frac{m}{4\pi}\ T(p,p';E)\rightarrow -\frac{m}{4\pi}\biggl[-{\cal N} g(p^2)\ g({p'}^2)\Delta(E)\biggr].
\label{Fullfofp}
\end{equation}
We then obtain the on-shell elastic scattering amplitude, ${\rm f}(p)$, and use 
\begin{equation}
{\rm f}(p)=f_2(p,p;E=p^2/m)=\frac{1}{p\cot\delta- ip},
\label{fofp}
\end{equation}
to determine
\begin{equation}
p \cot\delta=-\frac{\beta\gamma(2\beta+\gamma)}{2(\beta+\gamma)^2}+\frac{3\beta^2+2\beta\gamma+\gamma^2}{2\beta(\beta+\gamma)^2} p^2+\frac{1}{2\beta(\beta+\gamma)^2} p^4.
\label{pcotdel}
\end{equation}
Comparison with the Effective Range Expansion (ERE) 
\begin{equation}
p \cot\delta=-\frac{1}{a}+\frac{r_0}{2} p^2+v_4 p^4\dots,
\label{ERE}
\end{equation}
allows us to identify
\begin{equation}
a=\frac{2(1+\gamma/\beta)^2}{\gamma(2+\gamma/\beta)},\quad r_0=\frac{3+2\gamma/\beta+\gamma^2/\beta^2}{\beta(1+\gamma/\beta)^2}\quad {\rm and}\quad v_4=\frac{1}{2\beta^3(1+\gamma/\beta)^2} .
\label{EREconsts}
\end{equation}
Note that the ERE for the dipole form factor terminates at ${\cal O}(p^4)$.  To constrain the parameters, we fit our model to the phase shifts and binding energy as determined by the Nijmegen II (NIJMII) potential~\cite{Nijmegen:NNOnline} which gives $\gamma=0.231607$ fm$^{-1}$. Fitting to the NIJMII $^3$S$_1$ phase shifts for $p_{cm}^2< 0.85$ fm$^{-2}$, we determine $\beta=1.3940$ fm$^{-1}$.  Note that $\beta/\gamma=6.019$ in keeping with the expectation for the Hulthen wave function mentioned above.
We then {\it predict}
\begin{equation}
a=5.421\ (5.403)\ {\rm fm},\quad r_0=1.772\ (1.749)\ {\rm fm}\quad {\rm and}\quad v_4=0.1357\ (0.163)\ {\rm fm}^3,
\label{EREfit}
\end{equation}
which are in good agreement with{\it; e.g.}, Babenko and Petrov \cite{Babenko:2005qp} whose values appear above in the parentheses. This level of agreement is gratifying considering the simplicity of the approach. (We observe that a similar level of agreement is found upon assuming a Gaussian form factor although the ERE does not terminate at finite order in this case.) Note also that the value of $\beta$ is roughly two pion masses which is what we would expect qualitatively for this ``cutoff" parameter; {\it i.e.}, $\beta$ is ``natural''~\cite{Lepage:1990}. Finally we observe that none of the quantities in Eq.~\ref{EREconsts} is ill-behaved as $\beta\rightarrow\infty$; {\it i.e.}, as the potential becomes a simple contact ($\delta$-function) interaction. In this limit, $a\rightarrow1/\gamma$, $r_0\sim\beta^{-1}\rightarrow 0$, and $v_4\sim \beta^{-3}\rightarrow 0$ as expected.  We also emphasize that the ERE is fixed by the {\it on-shell} elastic amplitude, Eq.~\ref{fofp}, whereas, as we shall shortly see, the fully off-shell amplitude of Eq.~\ref{Fullfofp} is  -- in principle -- required in the 3-body Faddeev equations.

\section{3-Body Faddeev Equations using the Unitary Pole Approximation}
\label{Sec:FF}

Having explicit forms for the off-shell 2-body $T$-matrices allows one to calculate the 3-body scattering amplitude easily. The use of separable potentials in 3-body applications has a long and storied history to which we cannot do full justice here. Some recent work which has made use of separable potentials in 3-body calculations can be found in Refs.~\cite{Osman:1979,Haidenbauer:1986hk,Koike:1990}. Following Waston and Nuttall~\cite{Watson:1967} but using our sign conventions, the $nd$ $^4S_{3/2}$ $T$-matrix for the $\ell^{\rm th}$ partial wave is given by
\begin{equation}
{\cal X}_\ell(p, p';E)= 2 {\cal Z}_\ell(p, p';E)-\frac{1}{2 \pi^2} \int_0^\infty ~q^2 dq\ 2 {\cal Z}_\ell(p, q;E)~\Delta(E-\frac{3 q^2}{4 m})~{\cal X}_\ell(q, p';E), 
\label{X}
\end{equation}
where $\Delta(E)$ is given by Eq.~\ref{DeltaUPA}, identified with the deuteron (or {\it dimer}) propagator, and $\cal Z_\ell$ is the particle-exchange (``ping-pong") amplitude depicted diagrammatically in Fig.~\ref{ZDiag},
\begin{equation}
{\cal Z}_\ell(p, p';E)=-\Lambda_{IS} \frac{m}{2} \int_{-1}^{1}dx \frac{\psi_B(|{\bf p +p'/2}|)\psi_B(|{\bf p' +p/2}|)}{p^2+p'^2-m E^+ +p p' x}P_\ell(x),
\label{Z}
\end{equation}
where $P_\ell(x)$ is a Legendre polynomial and $\Lambda_{IS}$ is the isospin-spin structure factor which is unity for spinless bosons.  For the $nd$ quartet case,  
\begin{equation}
\Lambda_{IS}=U\biggl(\frac{1}{2},\frac{1}{2},\frac{3}{2},\frac{1}{2};1,1\biggr)\times U\biggl(\frac{1}{2},\frac{1}{2},\frac{1}{2},\frac{1}{2};0,0\biggr)=\biggl(1\biggr)\times\biggl(-\frac{1}{2}\biggr),
\label{SI}
\end{equation}
where the $U$'s are the unitary Racah recoupling coefficients. Note that the form factors for the 2-body bound state, $\psi_B$, appear in ${\cal Z}$. 

We can define the (fully off-shell) 3-body elastic scattering amplitude for the $\ell^{\rm th}$ partial wave by including the wavefunction normalization factor, Eq.~\ref{ZUPA}, as
\begin{equation}
f_\ell(p,p';E)= -\frac{m}{3\pi}Z_{B} {\cal X}_\ell(p,p';E).
\label{f3Def}
\end{equation}  
Then the on-shell 3-body partial wave scattering amplitude is in turn given by,
\begin{equation}
 f_\ell(p)=f_\ell\bigl(p,p;E=(3 p^2/4-\gamma^2)/m\bigr). 
\label{f3}
\end{equation}
Using Eq.~\ref{X}, the Faddeev equation for $f_\ell$ becomes:
\begin{equation}
f_\ell(p,p';E)=f_\ell^{Born}(p,p';E)+\int_0^\infty dq\ q^2 K_\ell(p,q;E) f_\ell(p,p';E).
\label{f3Eqn}
\end{equation}
For our form factor model we have
\begin{equation}
f_\ell^{Born}(p,p';E)=\frac{8\Lambda_{IS}\gamma}{3}(1+\gamma/\beta)^{3}\   {\cal Z}_{\ell}^{(0)}( p, p';E),
\label{fBornDef}
\end{equation}
with the reduced ${\cal Z}^{(0)}_\ell$ defined by
\begin{eqnarray}
{\cal Z}^{(0)}_{\ell}(p,q;E)&=& \int_{-1}^{+1} dx\ \frac{g(p^2/4+q^2+pqx)\ g(p^2+q^2/4+pqx)}{p^2+q^2-mE+pqx} P_\ell(x).
\label{Z0Def}
\end{eqnarray}
For $\ell=0$ specifically, using Eq.~\ref{DipoleFF} for $g(p)$, we have
\begin{eqnarray}
{\cal Z}^{(0)}_0(p,q;E)&=&\frac{4\beta^4}{3 p q (p^2-q^2)} \nonumber \\
& \times&\ \biggl[\frac{-1}{\beta^2+mE-3 q^2/4}  \ln \biggl(\frac{p^2+q^2-mE+pq}{p^2+q^2-mE-pq}\cdot \frac{\beta^2+p^2+q^2/4-pq}{\beta^2+p^2+q^2/4+pq}\biggr) \nonumber \\
&+& \quad \frac{1}{\beta^2+mE-3 p^2/4}  \ln \biggl(\frac{p^2+q^2-mE+pq}{p^2+q^2-mE-pq}\cdot \frac{\beta^2+p^2/4+q^2-pq}{\beta^2+p^2/4+q^2+pq}\biggr)\bigg].
\label{Z0swaveDef}
\end{eqnarray}
The {\it kernel}, $K_\ell$, is defined as
\begin{eqnarray}
K_\ell(p,q,E)&=& -\frac{1}{2\pi^2}\cdot  2{\cal Z}_\ell(p,q,E)\cdot \Delta(E-\frac{3q^2}{4 m}) \nonumber \\
&=&\frac{4\Lambda_{IS}}{\pi} \frac{(1+\gamma/\beta)^2}{\beta}\ {\cal Z}^{(0)}_{\ell}(p,q;E)\  \Biggl[1-\biggl(\frac{\beta+\gamma}{\beta+\sqrt{3 q^2/4-m E}}\biggr)^2\Biggr]^{-1}.
\label{KernelFF}
\end{eqnarray}
It is important to note that the kernel contains the three dynamical elements of the fully off-shell elastic scattering amplitude appearing in Eq.~\ref{Fullfofp}, namely the two form factors which reside in ${\cal Z}_\ell^{(0)}$ as given in Eq.~\ref{Z0Def} and the dimer propagator, $\Delta(E-3q^2/4 m)$. The arguments of these functions do not in general satisfy the on-shell condition. This dependence on the full off-shell structure of the 2-body amplitude is a well known characteristic of the 3-body Faddeev equations. As we shall soon see, in the standard ET approach, the form factors are -- in effect -- set equal to one and only the {\it on-shell} 2-body amplitude, Eq.~\ref{fofp}, enters.

Using Eqs.~\ref{fBornDef} and ~\ref{KernelFF}, we solve Eq.~\ref{f3Eqn} directly to find the 3-body elastic scattering amplitudes and phase shifts for the model.  (We employ the method of Hetherington and Schick~\cite{Hetherington:1965} to avoid ill-behaved integrands in the homogeneous term for energies above threshold.) Since $1/\beta$ is roughly the range of the interaction, we can study the effect of the finite range explicitly.  For this purpose we define Leading Order (FF-LO), Next-to-Leading Order (FF-NLO) and Next-to-Next-to-Leading Order (FF-NNLO) expressions by expanding $f_\ell^{Born}$ and $K_\ell$ on the rhs of Eq.~\ref{f3Eqn} in powers of $1/\beta$. The LO expression corresponds to retaining only terms independent of $1/\beta$; this obviously coincides with taking the $\beta\rightarrow\infty$ limit which corresponds to the simplest 2-body contact interaction. 
In particular the $\beta\rightarrow\infty$ limit of ${\cal Z}_0^{(0)}$ gives,
\begin{equation}
{\cal Z}_0^{(0)}(p,q,E)\ \stackrel{\longrightarrow}{_{_{\beta\rightarrow\infty}}}\  \frac{1}{pq}  \ln \biggl( \frac{p^2+q^2-mE+pq}{p^2+q^2-mE-pq}\biggr),
\label{ZLO}
\end{equation}
a factor that is ubiquitous in ET-ERE formulations. It is gratifying that this limit can be taken straightforwardly without relying on any particular prescription for renormalization of divergent integrals.  We note that the $\beta\rightarrow\infty$ form of ${\cal Z}_0^{(0)}$ is the only form which appears in the ET-ERE 3-body equations {\it at any order}! (See, for example, Ref.~\cite{Platter:2006ev}, specifically the discussion following their Eq.~34.) In contrast, we find, upon expanding ${\cal Z}_0^{(0)}(p,q,E)$, 
\begin{eqnarray}
{\cal Z}_0^{(0)}(p,q;E)&=& \frac{1}{pq}  \ln \biggl(\frac{p^2+q^2-mE+pq}{p^2+q^2-mE-pq}\biggr)
\times\biggl[ 1+\frac{3(p^2+q^2)-8mE}{4\beta^2}\biggr]-\frac{1}{4\beta^2}+{\cal O}\biggl(\frac{1}{\beta^4}\biggr),
\label{Z0exp}
\end{eqnarray}
which implies this quantity is modified from its $\beta\rightarrow\infty$ form starting at NNLO. We will comment on the ``extra" terms in the discussion to follow.

We next give explicit formulas for the $s$-wave amplitude; the subscript  $\ell=0$ will be implicit henceforth. 
Expanding the $s$-wave Born term, Eq.~\ref{fBornDef}, to second order, using Eq.~\ref{Z0swaveDef}, we find,
\begin{eqnarray}
f_{Born}&=& \frac{8\Lambda_{IS}\gamma}{3} \Biggl\{ \frac{1}{pq}  \ln \biggl(\frac{p^2+q^2-mE+pq}{p^2+q^2-mE-pq}\biggr)\nonumber \\
&\times&\biggl[ 1+3\frac{\gamma}{\beta}+\frac{1}{4\beta^2}\bigl(3p^2+3q^2+12\gamma^2- 8 mE\bigr)\biggr]-\frac{4}{\beta^2}\Biggr\}
+{\cal O}\biggl(\frac{1}{\beta^3}\biggr).
\label{f3Exp}
\end{eqnarray}
In similar fashion we expand the $s$-wave kernel, Eq.~\ref{KernelFF}, to find,
\begin{eqnarray}
K(p,q;E)&=& \frac{2\Lambda_{IS}}{\pi}\ \frac{1}{-\gamma+\sqrt{3 q^2/4-mE}} \Biggl\{
\frac{1}{pq}  \ln \biggl(\frac{p^2+q^2-mE+pq}{p^2+q^2-mE-pq}\biggr)\nonumber\\
&& \quad\biggl[ 1+\frac{3}{2\beta}(\gamma+\sqrt{3 q^2/4-mE})\nonumber \\
&&\quad\quad +\frac{1}{16\beta^2}\bigl(12 p^2+15 q^2+4\gamma^2-36 m E+40\gamma \sqrt{3 q^2/4-mE}\bigr)\biggr]\nonumber\\
 && \quad\quad\quad-\frac{4}{\beta^2}\Biggr\}
 +{\cal O}(1/\beta^3).
\label{KernelExp}
\end{eqnarray}
Similar expressions can be obtained for the higher partial waves.

We immediately see that the constant term, $-4/\beta^2$, in Eqs.~\ref{f3Exp} and \ref{KernelExp} will give a contribution at NNLO which will have no counterpart in the ET-ERE NNLO expressions (see Eq.~\ref{Z0exp}).  It is interesting to note that the form of this term is identical to the lowest order {\it 3-body} contact interaction invoked in the ET-ERE formulation. Upon reflection, the appearance of such a term is not surprising. We know that, for the 3-boson case ($\Lambda_{IS}=1$), the 3-body amplitude shows a log-periodic dependence on the upper limit of the $q$ integration in the homogeneous term~\cite{Bedaque:1998km,Griesshammer:2004pe}. Regularization is accomplished by introducing a simple 3-body contact interaction like the $-4/\beta^2$ term just mentioned. In the UPA version of the 3-boson problem, the form factor regulates the integral so no log-periodic dependence on the integration limit is found. Therefore, when ET-ERE formulations add 3-body terms, at least in part, they are compensating for additional 2-body finite range effects not accounted for in that approach.

Combining Eqs.~\ref{f3Exp} and \ref{KernelExp}, we find the Faddeev equations for our model at the various orders in the range of the interaction, $1/\beta$. First, at leading order, we find
\begin{eqnarray}
f_{LO}(p,p';E)&=&\frac{8\Lambda_{IS}\gamma}{3}\cdot \frac{1}{pp'}  \ln \biggl(\frac{p^2+{p'}^2-mE+pp'}{p^2+{p'}^2-mE-pp'}\biggr) \nonumber \\ 
&&+\frac{2\Lambda_{IS}}{\pi}\int_0^\infty q^2 dq\ \frac{1}{pq}  \ln \biggl(\frac{p^2+q^2-mE+pq}{p^2+q^2-mE-pq}\biggr)\frac{f_{LO}(q,p';E)}{-\gamma+\sqrt{3 q^2/4-mE}},
\label{f3LO}
\end{eqnarray}
which is, not surprisingly, identical to the ET-ERE equation at LO which will appear below (hence no notational distinction between models is needed at LO).  We again emphasize that this expression corresponds to the limit $\beta\rightarrow\infty$ which implies a simple 2-body contact interaction and that passing to this limit is trivial in the present formulation.With $\Lambda_{IS}\rightarrow 1$, this expression corresponds to the Skornyakov--Ter-Martirosyan (STM) equation for three identical spinless bosons~\cite{STM:1956,Danilov:1961}, and identical to the ET-ERE LO Faddeev equation~\cite{Bedaque:1998kg,Bedaque:1998km,Bedaque:1999vb,Bedaque:2002yg} (and many more; see,{\it e.g.}, Refs.\cite{Braaten:2004rn,Braaten:2006vd} for extensive reviews of this subject).   

Similarly, the NLO Faddeev equation for the form factor model is readily determined to be
\begin{eqnarray}
f^{FF}_{NLO}(p,p';E)&=&\frac{8\Lambda_{IS}\gamma}{3}\cdot \frac{1}{pp'}  \ln \biggl(\frac{p^2+{p'}^2-mE+pp'}{p^2+{p'}^2-mE-pp'}\biggr) \biggl[1+3\frac{\gamma}{\beta}\biggr]\nonumber \\ 
&&+\frac{2\Lambda_{IS}}{\pi}\int_0^\infty q^2 dq\ \frac{1}{pq}  \ln \biggl(\frac{p^2+q^2-mE+pq}{p^2+q^2-mE-pq}\biggr) 
\biggl[1+\frac{3}{2\beta}(\gamma+\sqrt{3 q^2/4-mE})\biggr]\nonumber \\ 
&&\qquad\qquad\frac {f^{FF}_{NLO}(q,p';E)}{-\gamma+\sqrt{3 q^2/4-mE}}.
\label{f3FFNLO}
\end{eqnarray}
Likewise for the FF-NNLO Faddeev equation, we find
\begin{eqnarray}
f^{FF}_{NNLO}(p,p';E)&=&\frac{8\Lambda_{IS}\gamma}{3}
\Biggl\{ \frac{1}{pp'}  \ln \biggl(\frac{p^2+{p'}^2-mE+pp'}{p^2+{p'}^2-mE-pp'}\biggr)\biggl[ 1+3\frac{\gamma}{\beta}+3\frac{\gamma^2}{\beta^2}+\frac{1}{4\beta^2}\bigl(3p^2+3{p'}^2- 8 mE\bigr)\biggr]\nonumber \\
&&\qquad\qquad -\frac{4}{\beta^2}\Biggr\} \nonumber \\  
&+&\frac{2\Lambda_{IS}}{\pi}\int_0^\infty q^2 dq\ \biggl\{\frac{1}{pq}  \ln \biggl(\frac{p^2+q^2-mE+pq}{p^2+q^2-mE-pq}\biggr)
\nonumber \\ &&\qquad\qquad\biggl[ 1+\frac{3}{2\beta}(\gamma+\sqrt{3 q^2/4-mE})+\biggl(\frac{3}{2\beta}(\gamma+\sqrt{3 q^2/4-mE}) \biggr)^2  \nonumber \\
&&\quad\quad +\frac{1}{4\beta^2}\bigl(3 p^2-3 q^2-8 \gamma(\gamma+\sqrt{3 q^2/4-mE}\ )\bigr)\biggr]-\frac{4}{\beta^2}\biggr\}\nonumber \\ &&\qquad\qquad\times\frac{f^{FF}_{NNLO}(q,p';E)}{-\gamma+\sqrt{3 q^2/4-mE}},
\label{f3FFNNLO}
\end{eqnarray}
where the first three terms in square brackets in the kernel are organized to resemble the corresponding expression for the ET-ERE to be considered next.

\section{3-Body ET-ERE Faddeev Equations}
\label{Sec:ET-ERE}

While the finite range expansion for our form factor model is straightforward, extending ET-ERE formulations to include corrections due to the finite range of the 2-body interaction in 3-body applications is not trivial and has been the subject of considerable attention in the literature; see Refs.~\cite{Griesshammer:2004pe,Platter:2006ev} for extensive discussions of this treatment and for additional related references. 

To emphasize the relation between our FF calculations as formulated in the preceding Section and the ET-ERE treatment, we obtain the latter by beginning with Eq.~\ref{X}, our full FF 3-body Faddeev equation for the amplitude ${\cal X}$ which we re-state as:
\begin{equation}
{\cal X}_\ell(p, p';E)= 2 {\cal Z}_\ell(p, p';E)-\frac{1}{2 \pi^2} \int_0^\infty ~q^2 dq\ 2 {\cal Z}_\ell(p, q;E)~\Delta\biggl(E-\frac{3 q^2}{4 m}\biggr)~{\cal X}_\ell(q, p';E), 
\nonumber
\end{equation}
where $\Delta(E)$ is is given by Eq.~\ref{DeltaE} and can be identified with the deuteron (or {\it dimer}) propagator and where, for $s$-wave scattering and assuming $\psi_B$ is real, ${\cal Z}_0$ 
is given by Eq.~\ref{Z} as
\begin{eqnarray}
{\cal Z}_0(p, p';E)&=&-\Lambda_{IS}\frac{m}{2} \int_{-1}^{1}dx \frac{\psi_B(|{\bf p +p'/2}|^2)\ \psi_B(|{\bf p' +p/2}|^2)}{p^2+p'^2-m E +p p' x} \nonumber \\
&=&-\Lambda_{IS}\frac{m}{2} {\cal N} \int_{-1}^{1}dx \frac{g(|{\bf p +p'/2}|^2)\ g(|{\bf p' +p/2}|^2)}{p^2+p'^2-m E +p p' x}\nonumber \\
&=&-\Lambda_{IS}\frac{m}{2} {\cal N}\ {\cal Z}^{(0)}_0(p,p';E),
\label{ZLO1}
\end{eqnarray}
where ${\cal N}$ is the normalization factor given in Eq.~\ref{DipoleNorm} and $g$ is the form factor (see Eq.~\ref{DipoleFF}).

We proceed toward the corresponding ET-ERE expression by first examining the homogenous term which we can write as
\begin{equation}
\frac{m\Lambda_{IS}\cal N}{2 \pi^2} \int_0^\infty q^2 dq\int_{-1}^{+1} dx\ \frac{g(|{\bf p +q/2}|^2)\ g(|{\bf q +p/2}|^2)}{p^2+q^2-m E +p qx} 
\cdot\Delta(E-\frac{3 q^2}{4 m})\cdot {\cal X}(q, p';E),
\label{Homo1}
\end{equation}
where the $\ell=0$ subscript has been suppressed. 
The next step is not physically justified but amounts to a prescription which yields the ET-ERE expressions. The arguments of the form factors are set to $mE-3 q^2/4$ in which case they no longer depend on the angular variable $x$. Thus the angular integral can be performed in the usual way and the homogeneous term becomes
\begin{equation}
\frac{2\Lambda_{IS}}{\pi} \int_0^\infty q^2 dq\ \frac{1}{p q}\  \ln \biggl(\frac{p^2+q^2-mE+pq}{p^2+q^2-mE-pq}\biggr)\ 
\biggl[\frac{m{\cal N}}{4\pi}\cdot g\bigl(mE-3 q^2/4\bigr)^2\cdot \Delta(E-\frac{3 q^2}{4 m})\biggr]\ {\cal X}(q, p';E).
\label{Homo2}
\end{equation}
Now, according to Eqs.~\ref{LSeq2},~\ref{DeltaDef} and~\ref{Fullfofp}, the quantity in the square brackets is precisely the on-shell 2-body elastic scattering amplitude, ${\rm f}(p)$, evaluated at $p \leftrightarrow \sqrt{mE-3 q^2/4}$. Now using, Eq.~\ref{fofp} and the Effective Range Expansion, Eq.~\ref{ERE}, we can rewrite the homogeneous term as
\begin{eqnarray}
&&\frac{2}{\pi} \int_0^\infty q^2 dq\ \frac{1}{p q}\  \ln \biggl(\frac{p^2+q^2-mE+pq}{p^2+q^2-mE-pq}\biggr)\cdot {\rm f}^{-1}\bigl(p=\sqrt{mE-3q^2/4}\bigr)\cdot {\cal X}(q, p';E)\nonumber\\
&=&\frac{2}{\pi} \int_0^\infty q^2 dq\ \frac{1}{p q}\  \ln \biggl(\frac{p^2+q^2-mE+pq}{p^2+q^2-mE-pq}\biggr)\ 
\frac{{\cal X}(q, p';E)}{-\frac{1}{a}+\frac{r_0}{2}(m E-3q^2/4)+v_4(m E-3q^2/4)^2\dots+\sqrt{3q^2/4-m E}}.\nonumber\\
\label{Homo3}
\end{eqnarray}

Next we look at the Born term in the equation for ${\cal X}$. Using the expression for ${\cal Z}_\ell(p,p';E)$ appearing in Eq.~\ref{Z}, the Born term in Eq.~\ref{X} for our full FF 3-body Faddeev equation for $\ell=0$ is,
\begin{equation}
2{\cal Z}(p, p';E)=-m \Lambda_{IS} \int_{-1}^{1}dx \frac{\psi_B(|{\bf p +p'/2}|)\psi_B(|{\bf p' +p/2}|)}{p^2+p'^2-m E^+ +p p' x},\nonumber
\end{equation}
where here and henceforth the $\ell=0$ subscript is suppressed. 
To make contact with the ET-ERE approach, we {\it ignore} the momentum dependence of the $\psi_B$'s, perform the integral over $x$ and rewrite the Born term as
\begin{equation}
2{\cal Z}(p, p';E)\rightarrow-m\ \Lambda_{IS}\ {\cal N}\cdot \frac{1}{p p'}\  \ln \biggl(\frac{p^2+{p'}^2-mE+p p'}{p^2+{p'}^2-mE-p p'}\biggr).
\label{ZETDef}
\end{equation}
where $ {\cal N}$ is the form factor normalization (see Eq.~\ref{DipoleNorm}). We remind the reader that the assumption made above in deriving this simplified Born term -- ignoring the momentum dependence at the vertices of the ``ping-pong" diagram -- is always made implicitly and its validity is never checked. We next find the Born contribution to the 3-body elastic scattering amplitude, $f$, by making use of the relation between $f$ and ${\cal X}$ 
 as given in Eq.~\ref{f3Def}:
\begin{eqnarray}
f^{ET-ERE}_{Born}=\frac{m^2\Lambda_{IS}}{3\pi}\ Z^{ERE}\ \frac{1}{p p'}\  \ln \biggl(\frac{p^2+{p'}^2-mE+p p'}{p^2+{p'}^2-mE-p p'}\biggr),
\label{f3ETBorn}
\end{eqnarray}
where $Z^{ERE}\leftrightarrow Z_B {\cal N}$ is the ET-ERE wave function renormalization to be discussed shortly.

Then, using the expression in Eq.~\ref{f3ETBorn} for the Born term and the expression for the homogeneous term in Eq.~\ref{Homo3}, the ET-ERE Faddeev equation for the 3-body scattering amplitude becomes
\begin{eqnarray}
f^{ET-ERE}&=&\frac{m^2\Lambda_{IS}}{3\pi}\ Z^{ERE}\ \frac{1}{p p'}\  \ln \biggl(\frac{p^2+{p'}^2-mE+p p'}{p^2+{p'}^2-mE-p p'}\biggr)\nonumber \\
&&\quad+ \frac{2}{\pi} \int_0^\infty q^2 dq\ \frac{1}{p q}\  \ln \biggl(\frac{p^2+q^2-mE+pq}{p^2+q^2-mE-pq}\biggr)\nonumber \\
&&\qquad\quad \times\frac{f^{ET-ERE}(q, p';E)}{-\frac{1}{a}+\frac{r_0}{2}(m E-3q^2/4)+v_4(m E-3q^2/4)^2\dots+\sqrt{3q^2/4-m E}}.
\label{f3ETFullNew}
\end{eqnarray}

It turns out in practice to be more convenient to use a form of ${\rm f}^{-1}(p)$ which is equivalent to using a version of the ERE based on an expansion of $p\cot\delta$ about $p^2=-\gamma^2$ rather than about $p^2=0$ as is implicit in Eq.~\ref{ERE}. Thus
\begin{equation}
{\rm f}^{-1}(p)\rightarrow-\gamma+\frac{r'_0}{2}(p^2+\gamma^2)+v'_4(p^2+\gamma^2)^2\dots -i p,
\label{fINVaboutGamma}
\end{equation}
where there is clearly a simple relation between the original ERE parameters in Eq.~\ref{ERE} and their primed counterparts in the last equation. (For example, we have $r_0=1.772$ fm but $r'_0=1.747$ fm based on our dipole form factor.) 
We will adopt this latter prescription to generate our ET-ERE NLO and N$^2$LO Faddeev equations from Eq.~\ref{f3LO}.  Other methods such as the  $Z$-expansion approach appear in the literature; we refer the reader to Refs.~\cite{Platter:2006ev} for details. In all ET-ERE cases we have examined, the various forms of the ERE expansion yield results which are qualitatively similar.
 
Using the form of ${\rm f}(p)$ given in Eq.~\ref{fINVaboutGamma}, our putative higher order ET-ERE s-wave Faddeev equation is then
\begin{eqnarray}
f^{ET-ERE}(p,p';E)&=&\frac{m^2\Lambda_{IS}}{3\pi}\ Z^{ERE}\ \cdot \frac{1}{pp'}  \ln \biggl(\frac{p^2+{p'}^2-mE+pp'}{p^2+{p'}^2-mE-pp'}\biggr) \nonumber \\ 
&&+\frac{2\Lambda_{IS}}{\pi}\int_0^\infty q^2 dq\ \frac{1}{pq}  \ln \biggl(\frac{p^2+q^2-mE+pq}{p^2+q^2-mE-pq}\biggr)f^{ET-ERE}(q,p';E)\nonumber \\ 
&&\qquad\qquad\biggl[-\gamma+\frac{r_0'}{2} (mE+\gamma^2-3 q^2/4)+{v'}_4 (mE+\gamma^2-3 q^2/4)^2 \nonumber \\ 
&&\qquad\qquad\qquad\qquad+\dots+\sqrt{3 q^2/4-mE}\biggr]^{-1} ,
\label{f3ETFull}
\end{eqnarray}
which apparently includes contributions from the ERE {\it to all orders}. However, as has been noted frequently in the literature~\cite{Braaten:2004rn,Griesshammer:2004pe}, use of this form of what amounts to the dimer propagator in 3-body Faddeev equations is problematic as it introduces spurious poles in the integrand of the homogeneous term which lead to unphysical effects, the most serious of which are imaginary phase shifts that are non-zero below the breakup threshold or that are negative. A nearly universally applied prescription for avoiding (some of) these pitfalls is to use (see, {\it e.g.}, Ref.\cite{Bedaque:2002yg} for one of the earliest discussions of this point) for ${\rm f}(p)$, instead of Eq.~\ref{fINVaboutGamma},
\begin{equation}
{\rm f}(p)\rightarrow\frac{1}{-\gamma-i p}\biggl[1+\sum_{n=1}^{N_{max}}\biggl(\frac{r_0'}{2}\cdot (\gamma-i p) \biggr)^n\ \biggr],
\label{fINVERE}
\end{equation}
which obviously only takes into account the ${\cal O}(p^2)$ term in the ERE but which preserves the pole structure of the LO  integrand (Eq.~\ref{f3LO}) in the homogeneous term.  We henceforth adopt the terminology frequently used in the literature to designate calculations employing Eq.~\ref{fINVERE} as NLO or N$^2$LO for $N_{max}=1$ or 2, respectively. We adopt these prescriptions to generate our ET-ERE NLO and N$^2$LO Faddeev equations by using the form of ${\rm f}(p)$ appearing in Eq.~\ref{fINVERE} in the 3-body Faddeev equation shown in Eq.~\ref{f3ETFull}. 

To find explicit ET-ERE 3-body equations, it remains to specify the wave function renormalization, $Z_{ERE}$. Following, {\it e.g.}, Ref.~\cite{Bedaque:1998kg,Bedaque:1998km,Griesshammer:2004pe} , we write
\begin{eqnarray}
Z_{ERE}^{-1}&=&-\frac{m}{4\pi}\ \frac{d\ }{dE}\bigg|_{E=-B} {\rm f}_{ERE}^{-1}(p=\sqrt{mE^+}).
\label{ZERE}
\end{eqnarray}
According to Eq.~\ref{fINVERE}, at LO, $ {\rm f}^{-1}_{ERE}(p=\sqrt{mE^+})\rightarrow-\gamma+\sqrt{-mE^+}$ which implies, with $B=\gamma^2/m$, 
\begin{eqnarray}
Z^{ERE}_{LO}&=&\frac{8\pi\gamma}{m^2}.
\label{ZERELO}
\end{eqnarray}
Similarly, we readily find
\begin{eqnarray}
Z^{ERE}_{NLO}&=&\frac{8\pi\gamma}{m^2}(1+r_0'\gamma),\quad\quad Z^{ERE}_{NNLO}=\frac{8\pi\gamma}{m^2}(1+r_0'\gamma+r_0'^2\gamma^2).
\label{ZERENNLO}
\end{eqnarray}

Thus, at NLO we find
\begin{eqnarray}
f^{ET-ERE}_{NLO}(p,p';E)&=&\frac{8\gamma\Lambda_{IS}(1+r_0'\gamma)}{3}\cdot \frac{1}{pp'}  \ln \biggl(\frac{p^2+{p'}^2-mE+pp'}{p^2+{p'}^2-mE-pp'}\biggr) \nonumber \\ 
&&+\frac{2\Lambda_{IS}}{\pi}\int_0^\infty q^2 dq\ \frac{1}{pq}  \ln \biggl(\frac{p^2+q^2-mE+pq}{p^2+q^2-mE-pq}\biggr)\frac{f^{ET-ERE}_{NLO}(q,p';E)}{-\gamma+\sqrt{3 q^2/4-mE}}\nonumber \\ 
&& \qquad\qquad\times\biggl[1+\frac{r_0'}{2}\bigl(\gamma+\sqrt{3 q^2/4-mE}\bigr)\biggl],
\label{f3ETNLO}
\end{eqnarray}
while at N$^2$LO
\begin{eqnarray}
f_{NNLO}^{ET-ERE}(p,p';E)&=&\frac{8\gamma\Lambda_{IS}(1+r'_0\gamma+r_0'^2\gamma^2)}{3}\cdot \frac{1}{pp'}  \ln \biggl(\frac{p^2+{p'}^2-mE+pp'}{p^2+{p'}^2-mE-pp'}\biggr) \nonumber \\ 
&&+\frac{2\Lambda_{IS}}{\pi}\int_0^\infty q^2 dq\ \frac{1}{pq}  \ln \biggl(\frac{p^2+q^2-mE+pq}{p^2+q^2-mE-pq}\biggr)\frac{f_{NNLO}^{ET-ERE}(q,p';E)}{-\gamma+\sqrt{3 q^2/4-mE}}\nonumber \\ 
&& \qquad\qquad\times\biggl[1+\frac{r_0'}{2}\bigl(\gamma+\sqrt{3 q^2/4-mE}\bigr)\nonumber \\ 
&&\qquad\qquad\qquad\qquad+\biggl(\frac{r_0'}{2}\bigl(\gamma+\sqrt{3 q^2/4-mE}\bigr)\biggr)^2\biggl].
\label{f3ETNNLO}
\end{eqnarray}

Comparing the FF and ET-ERE Faddeev equations at NLO, we find Eqs.~\ref{f3FFNLO} and \ref{f3ETNLO} would be identical if we could equate  $r'_0$ with $3/\beta$. This relation is found to be satisfied numerically with reasonable accuracy ($r'_0\simeq1.75$ fm vs  $3/\beta\simeq 2.16$ fm), but the difference is a real one since we must use the parameters as set by the 2-body phenomenology if we are to be consistent.  We then must conclude that incorporating finite range effects via our separable potential with a simple dipole form factor gives rise to contributions to 3-body scattering which cannot be extracted exclusively from the 2-body ERE.

Next we compare the FF and ET-ERE Faddeev equations at NNLO; {\it i.e.} Eqs.~\ref{f3FFNNLO} and \ref{f3ETNNLO}, where significant differences are apparent. The three terms in the square brackets in the homogeneous term of the ET-ERE Faddeev equation, Eq.~\ref{f3ETNNLO}, also appear in that position in the FF NNLO equation, Eq.~\ref{f3FFNNLO}. These terms can be traced back to the expansion employed in the ET-ERE for ${\rm f}(p)$ which is specified in Eq.~\ref{fINVERE}. However, extra terms at ${\cal O}(1/\beta^2)$ are present in the FF expression including the $-4/\beta^2$ term which resembles a 3-body contact interaction as discussed previously (see Eq.~\ref{Z0exp}). We conclude that, while the ET-ERE, as implemented employing Eq.~\ref{fINVERE} indeed avoids pathological pole structure of ${\rm f}(p)$ in the homogeneous term of the ET-ERE 3-body Faddeev equations, it ignores the form factors appearing in Fig.~\ref{ZDiag} and hence fails to account for the momentum dependence contained in the form factor vertices associated with the 2-body interaction. Consequences of these shortcomings will be examined in the next Section.

\section{Results}
\label{sec:results}

We now present calculations based on the expressions developed thus far. As discussed in Sect.~\ref{Sec:ERE}, our 2-body input is constrained by the Nijmegen II (NIJMII) $np$ phase shifts in the $^3S_1$ channel.   We first determine $p \cot\delta$ and $\gamma=0.231607$ fm$^{-1}$ from the NIJMII potential. We then adjust $\beta$ in Eq.~\ref{pcotdel} to give a fit to $p \cot\delta$ up to $p^2\ \sim 0.85$ fm$^{-2}$, finding $\beta=1.3940$ fm$^{-1}$. We then use the expressions in the previous section to determine s-wave phase shifts for the $^4S_{3/2}$ (quartet) channel in $nd$ elastic scattering. To provide a context for comparison, we also show 3-body phase shifts from full Faddeev calculations using realistic NN potentials, Refs.~\cite{Huber:1993,Kievsky:2004-737}. 

\begin{figure}[ht]
\vspace{0.50in}
\includegraphics[width=7in,angle=0]{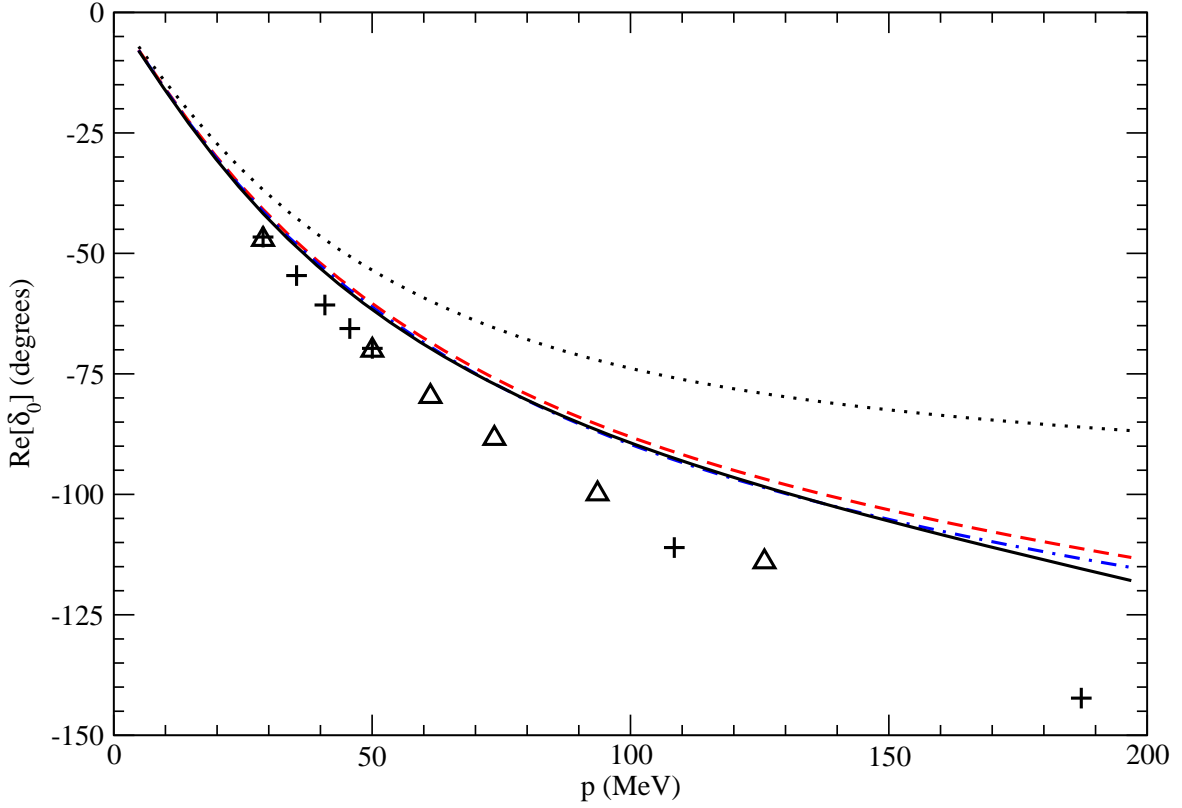} 
\caption{(Color online).
The real part of $L=0$ $nd$ quartet phase shifts for the artificial case, $\beta\rightarrow 2\beta$. The solid curve is the full FF calculation, Eqs.~\ref{f3Eqn}-\ref{KernelFF}, and the dotted curve is the LO calculation,  Eq.~\ref{f3LO}. The other curves are NLO calculations in the ET-ERE formulation (dash),  Eqs.~\ref{f3ETNLO}-\ref{f3ETNNLO} and FF-ERE formulation (dash-dot), Eqs.~\ref{f3FFNLO}-\ref{f3FFNNLO}.  The NNLO calculations are identical to the NLO indicating rapid convergence as expected for this case using an artificially short range parameter.  All FF calculations are based on a dipole form factor with with an original inverse range parameter of $\beta=1.394$ fm$^{-1}$ determined by fitting to $p\cot\delta$ as given by the Nijmegen-II $^3$S$_1$ phase shifts~\cite{Nijmegen:NNOnline}. In the ET-ERE calculations, multiplying the original $\beta$ by 2 implies an effective range parameter $r_0\sim 0.97$ fm.  Also shown are the full Faddeev calculations of Huber, et al. ~\cite{Huber:1993} (triangle) and Kievsky, et al.~\cite{Kievsky:2004-737} (plus).}
\label{FR_Re_S_low_r0}
\end{figure}

\begin{figure}[ht]
\vspace{0.50in}
\includegraphics[width=7in,angle=0]{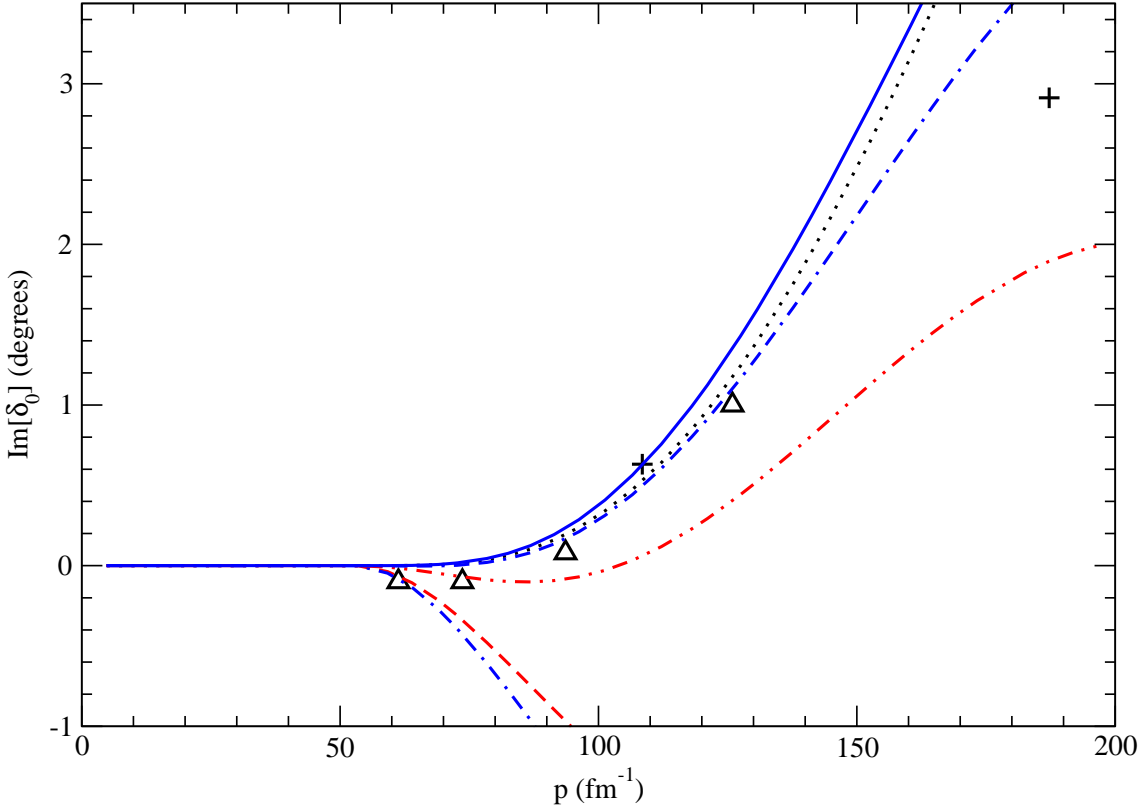} 
\caption{(Color online).
The imaginary part of $L=0$ $nd$ quartet phase shifts for the artificial case, $\beta\rightarrow 2\beta$. The solid curve is the full FF calculation, Eqs.~\ref{f3Eqn}-\ref{KernelFF}, and the dotted curve is the LO calculation,  Eq.~\ref{f3LO}. The other curves are NLO and NNLO calculations in the ET-ERE formulation (dash and dash-dot-dot, respectively),  Eqs.~\ref{f3ETNLO}-\ref{f3ETNNLO}, and FF-ERE formulation (dash-dot and dash-dash-dot, respectively), Eqs.~\ref{f3FFNLO}-\ref{f3FFNNLO}. All FF calculations are based on a dipole form factor with with an original inverse range parameter of $\beta=1.394$ fm$^{-1}$ determined by fitting to $p\cot\delta$ as given by the Nijmegen-II $^3$S$_1$ phase shifts~\cite{Nijmegen:NNOnline}. In the ET-ERE calculations, multiplying the original $\beta$ by 2 implies an effective range parameter $r_0\sim 0.97$ fm. Also shown are the full Faddeev calculations of Huber, et al. ~\cite{Huber:1993} (triangle) and Kievsky, et al.~\cite{Kievsky:2004-737} (plus).
 Note that breakup threshold is at $p=53$ MeV/c.}
\label{FR_Im_S_low_r0}
\end{figure}

For our first set of calculations we demonstrate the convergence of the finite range expansion by {\it artificially} letting $
\beta\rightarrow 2\times\beta\sim 2.79$ fm$^{-1}$ 
which means the 2-body effective range parameter, $r_0$,  decreases by roughly one half.   
The expansions of the 3-body Faddeev equations which give our Leading Order (LO), Next-to-Leading Order (NLO) and Next-to-Next-to-Leading Order (NNLO) expressions of the preceding Section -- which are expansions in $\gamma/\beta$ in the FF case or $\gamma r_0'$ for the ET-ERE case -- will converge rapidly since, eg,  $\gamma/\beta\simeq 
1/6\rightarrow 1/12$.   
Our calculations of the real and imaginary phase shifts for this artificial case appear in Figs.~\ref{FR_Re_S_low_r0} and ~\ref{FR_Im_S_low_r0}, respectively.  

For the real part of the phase shifts, Fig.~\ref{FR_Re_S_low_r0}, both the FF and ET-ERE calculations have converged at NLO and that there is little difference between the two approaches here. (Agreement with the full Faddeev calculations is understandably poor since we have artificially increased $\beta$.) Examination of the imaginary part of the phase shifts, Fig.~\ref{FR_Im_S_low_r0}, is a bit more revealing. Here we see some differences at the various orders of the expansion, with the NLO results lying below the NNLO and full FF values. (The fact that the LO calculations are in near agreement with the NNLO and full FF results is coincidental.) At each order (LO, NLO and NNLO), the FF and ET-ERE calculations are quite similar suggesting that, in this artificial case where convergence has been accelerated, the two approaches are nearly equivalent. Closer inspection reveals two interesting features. First, both NLO calculations yield slightly {\it negative} imaginary phase shifts from threshold up to about $p=110$ MeV/c. Though these negative imaginary phase shifts are small in magnitude, the are large compared with our numerical uncertainties and, from a physical point of view, they represent a serious failure as they are incompatible with unitarity. This failure is evidently corrected in this case by going to NNLO where the imaginary phase shifts are positive above threshold as we would expect. The second point of interest is that, at NNLO, the FF phase shifts are much closer to the full FF values than are the ET-ERE results. This is perhaps not surprising since the FF NNLO is based directly on the expansion the full FF in powers of $1/\beta$ and presumably includes, {\it e.g.}, some of the off-shell properties of the 2-body $T$-matrix which cannot be accounted for in the ET-ERE formulation.

\begin{figure}[ht]
\vspace{0.50in}
\includegraphics[width=7in,angle=0]{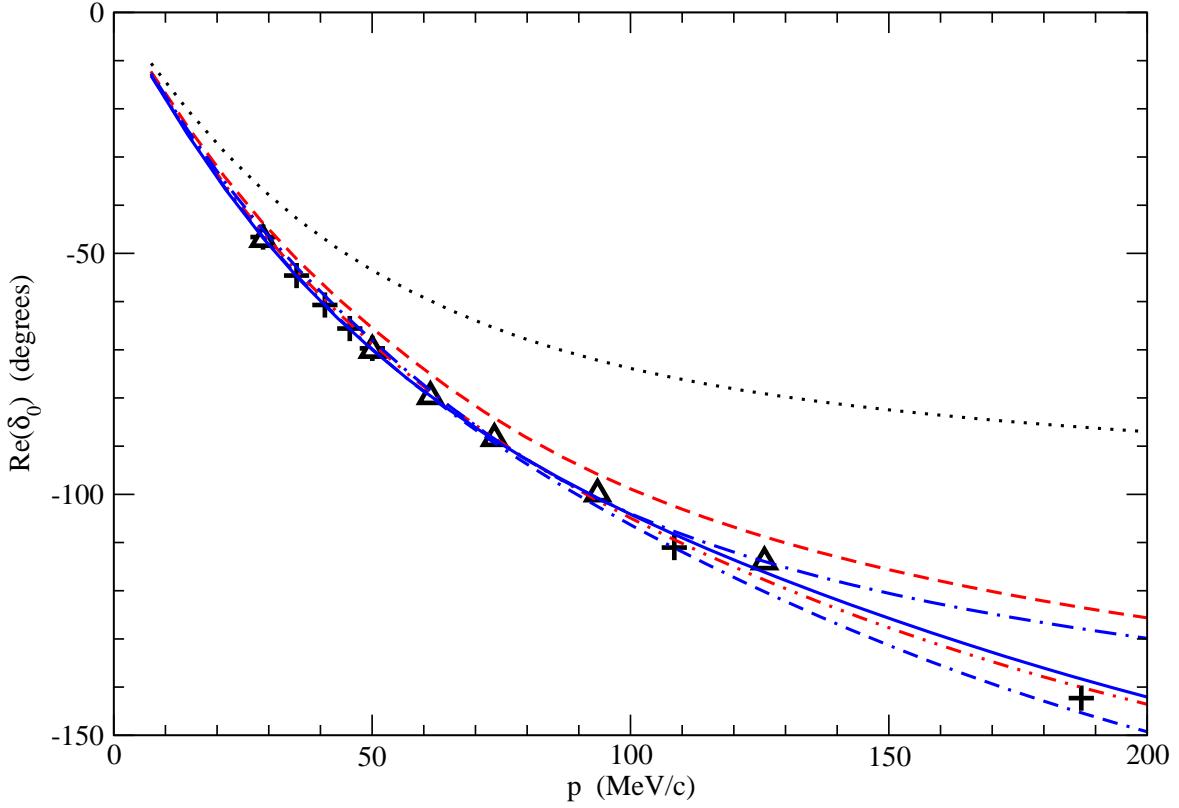}
\caption{(Color online).
The real part of $L=0$ $nd$ quartet phase shifts. The solid curve is the full FF calculation, Eqs.~\ref{f3Eqn}-\ref{KernelFF} and the dotted curve is the LO calculation, Eq.~\ref{f3LO}. The other curves are NLO and NNLO calculations in the ET-ERE (dash and dash-dot-dot, respectively),  Eqs.~\ref{f3ETNLO}-\ref{f3ETNNLO}, and FF-ERE (dash-dot and dash-dash-dot, respectively), Eqs.~\ref{f3FFNLO}-\ref{f3FFNNLO}, formulations. All FF calculations are based on a dipole form factor with $\beta=1.394$ fm$^{-1}$ determined by fitting to $p\cot\delta$ as given by the Nijmegen-II $^3$S$_1$ phase shifts~\cite{Nijmegen:NNOnline}. In the ET-ERE calculations, this corresponds to an effective range parameter $r_0'=1.747$ fm. Also shown are the full Faddeev calculations of Huber, et al. ~\cite{Huber:1993} (triangle) and Kievsky, et al.~\cite{Kievsky:2004-737} (plus).}
\label{FR_Re_S}
\end{figure}

\begin{figure}[ht]
\vspace{0.50in}
\includegraphics[width=7in,angle=0]{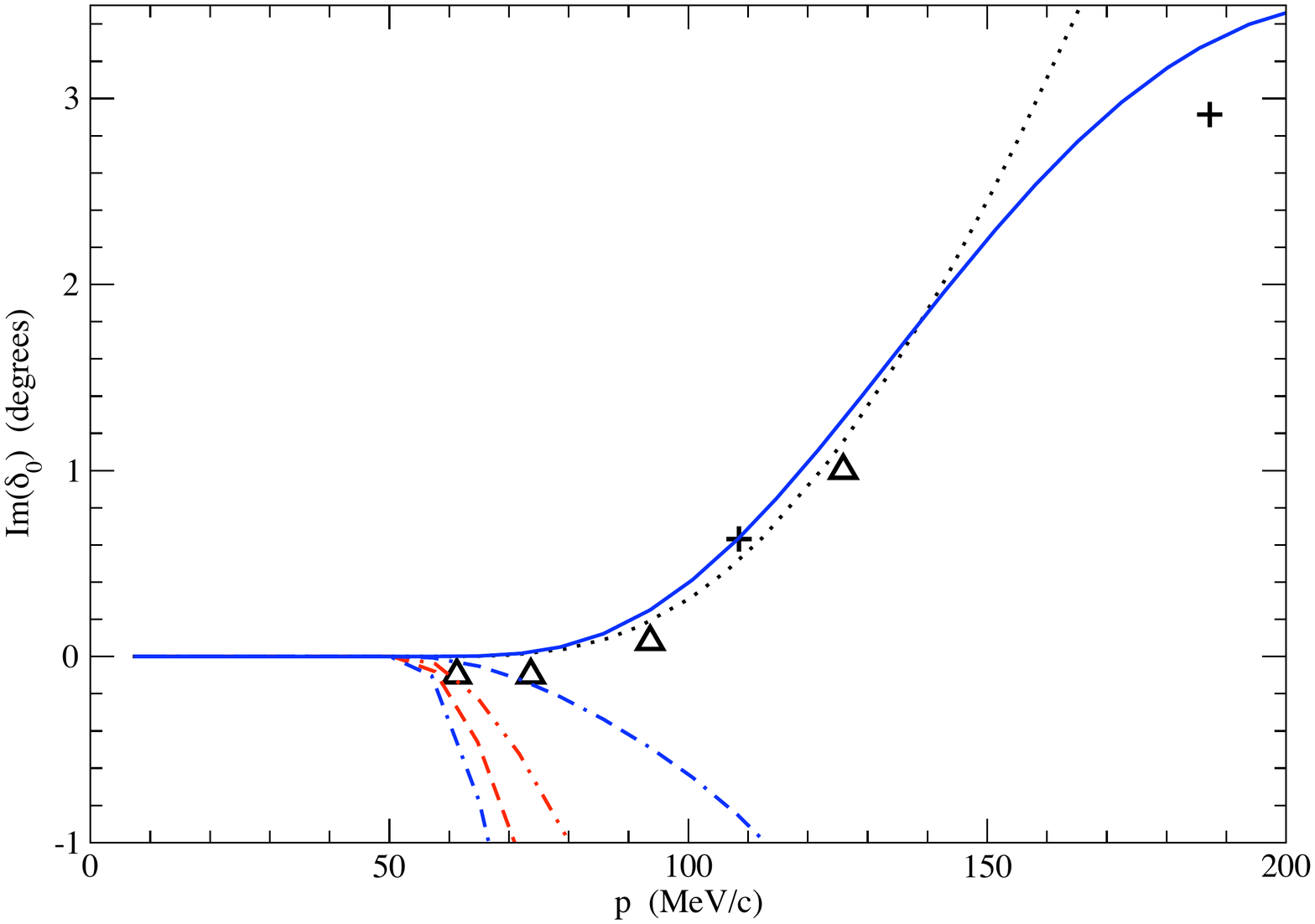}
\caption{(Color online).
         Same as for Fig.~\ref{FR_Re_S} but for the imaginary part of the  $nd$ quartet phase shifts. Note that breakup threshold is at $p=53$ MeV/c.}
\label{FR_Im_S}
\end{figure}

We now turn to realistic $nd$ calculations which employ the value of $\beta=1.3904$ fm$^{-1}$ fit to the NIJMII phase shifts. We first consider the quartet scattering length $^4a_3$, values for which are presented in Table~\ref{Table2}. Here our results for the various calculations under discussion are compared with the old experimental results of Dilg {\it et al.}~\cite{Dilg:1971} and the modern full Faddeev results of Wita{\l}a {\it et al.}~\cite{Witala:2003mr}. We focus on the latter result as it corresponds to the specific case of the NIJMII potential without electromagnetic interactions and hence most closely resembles our NN input. We see that the LO, NLO and NNLO calculations of both types steadily converge toward the Witala {\it et al.} value of 6.325 fm, but that the FF results appear to converge somewhat faster. 
Our full FF calculation agrees almost exactly with Wita{\l}a {\it et al.} value. Next we look at the real and imaginary s-wave phase shifts which appear in Figs.~\ref{FR_Re_S} and ~\ref{FR_Im_S}, respectively. For this realistic case, differences among the various calculations are apparent for both the real and imaginary phase shifts. For the real phase shifts, both the FF and ET-ERE calculations appear to have converged at NNLO, but the convergence of the former seems a bit faster. Both the FF NNLO and full FF calculations agree with the full Faddeev results quite well while the ET-ERE results are arguably just as good. 

Examination of the imaginary phase shifts in Fig.~\ref{FR_Im_S} leads to rather different conclusions. First, the failure in the form of negative imaginary phase shifts has become catastrophic. The NLO and NNLO calculations of both types yield large negative imaginary phase shifts over the entire momentum range covered. We have carried out higher order calculations (not shown) and find that the failure is corrected by $N^3LO$ and $N^4LO$.  Moreover, the FF NNLO calculation again seems to be appreciably closer to the full calculations than its ET-ERE counterpart. These findings suggest (i) that finite range expansions of either type should not be relied upon without carefully checking convergence properties and (ii) that expansions based on the on-shell information contained in the 2-body Effective Range Expansion may be inadequate considering the appearance of the full off-shell $T$-matrix in the original Faddeev formulation and in the FF treatment considered here. We also note that the full FF phase shifts are in excellent agreement with the full Faddeev calculations shown.

As an additional test of the finite range expansions, Table~\ref{Table1} shows calculations of the (center of mass) total, elastic, break-up, and back-angle cross sections for quartet nd scattering at $E_{lab}=10$ MeV, summing partial waves up to $\ell=8$ for each of the models under study.  For comparison purposes the table also shows the calculations of Koike and Taniguchi~\cite{Koike:1986}.  (These authors point out that the $nd$ elastic cross section at 180$^\circ$ is dominated by the quartet contribution so separation from the doublet is not required.) The ET-ERE and FF NNLO calculations are typically close to one another, but each falls appreciably below both the full FF calculation and the realistic calculations of Koike and Taniguchi. The discrepancy of the NNLO calculations with the other results for the breakup cross section is traceable in part to the negative imaginary phase shifts of the former. It is gratifying to note that results from our simple FF model are consistently within 10\% of the Koike and Taniguchi values.

\begin{figure}[ht]
\vspace{0.50in}
\includegraphics[width=7in,angle=0]{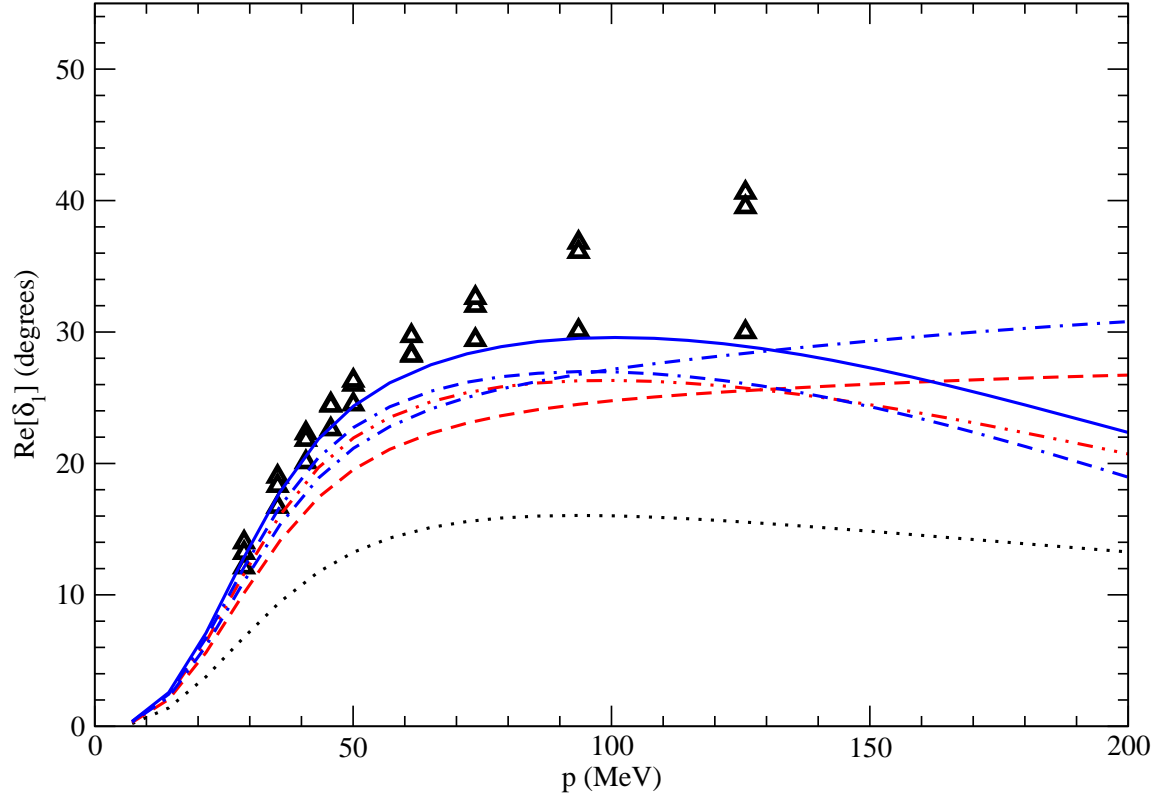}
\caption{
(Color online). Same as Fig.~\ref{FR_Re_S} except for the real part of $L=1$ $nd$ quartet phase shifts.}
\label{FR_Re_P}
\end{figure}

\begin{figure}[ht]
\vspace{0.50in}
\includegraphics[width=7in,angle=0]{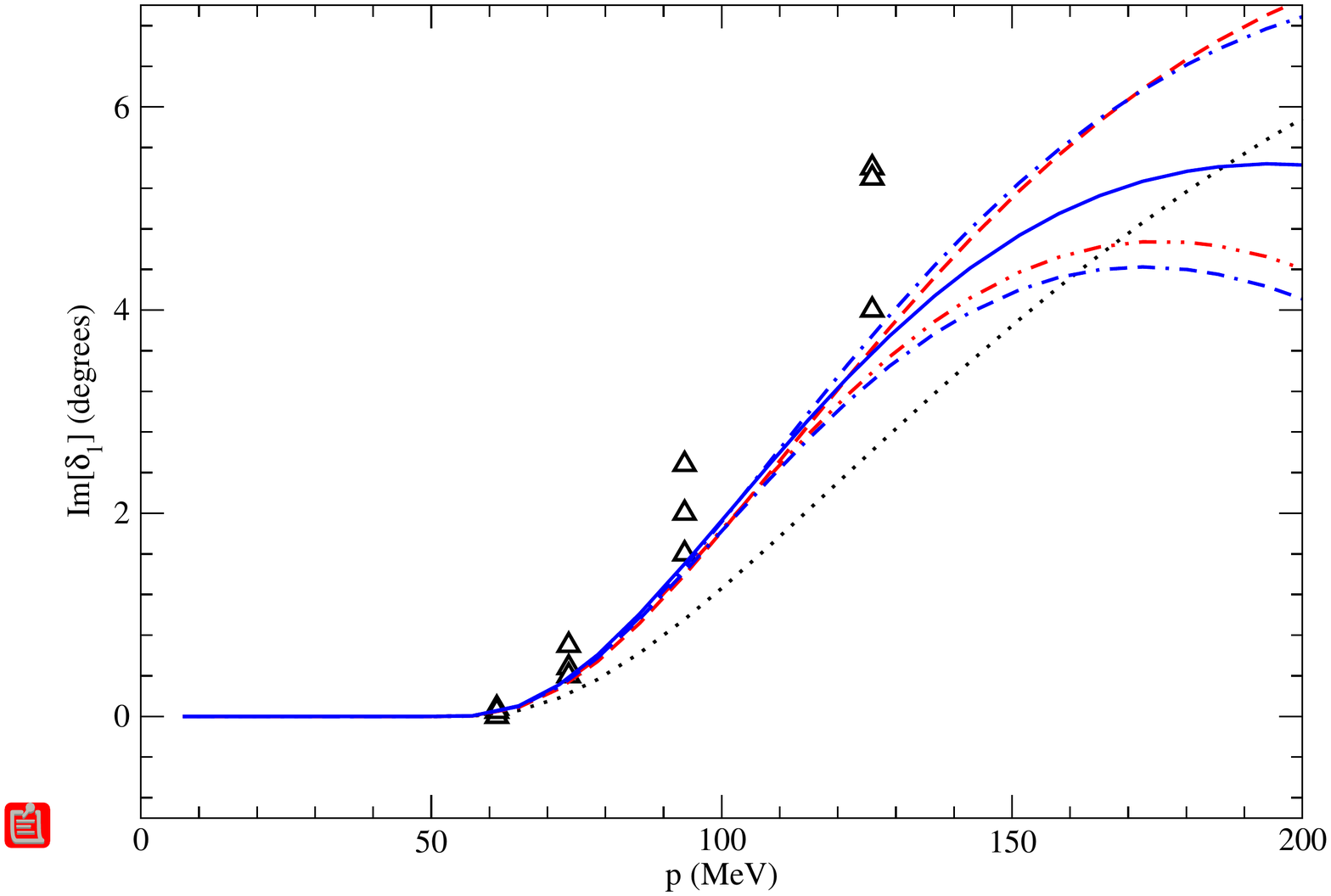}
\caption{
(Color online).  Same as for Fig.~\ref{FR_Re_S} except for the imaginary part of the $L=1$  $nd$ quartet phase shifts. Note that breakup threshold is at $p=53$ MeV/c.}
\label{FR_Im_P}
\end{figure}

We now turn to calculations of the $\ell=1$ $nd$ quartet phase shifts shown in Figures~\ref{FR_Re_P} and~\ref{FR_Im_P} and similarly show LO, NLO, and NNLO calculations for ET-ERE while taking the FF to N$^2$LO 
as well as the full form factor calculation and the full Faddeev calculations of \cite{Huber:1993}.

For the real part of $p$-wave quartet phase shift, Fig.~\ref{FR_Re_P}, we see that (i) the full form factor calculation is in good agreement with the lower range of the full Faddeev calculations, (ii) the ET-ERE NLO and FF NLO are essentially identical but differ at NNLO.  We have carried out higher order calculations (not shown) which oscillate about, but converge to, albeit slowly, the full FF result.

The situation is more dramatic for the imaginary part of the $p$-wave phase shift as shown in Fig.~\ref{FR_Im_P}.  Again the full form factor calculation is in fair agreement with the full Faddeev calculations. The LO calculations look quite good as do both the ET-ERE NLO and FF NLO calculations which are essentially identical. However, the ET-ERE NNLO and FF NNLO diverge from one another, apparently due to the extra terms in the latter (Eq.~\ref{f3FFNNLO}) which are not present in the former (ET-NNLO expression, Eq.~\ref{f3ETNNLO}).  Again, the higher order FF calculations oscillate about the final full form factor results. Generally, the calculations relying on an expansion in the range of the interaction display poor convergence properties while the full form factor results are well behaved and, indeed, give an excellent account of both the real and imaginary phase shifts examined here.

At this point, we address two sets of ET-ERE calculations which {\it do not} use the prescription implied by Eq.~\ref{fINVERE}. The first is the work of Bedaque and Griesshammer~\cite{Bedaque:1999vb} in which they present calculations of quartet $s$-wave phase shifts based on the ET approach. These constitute, to the best of our knowledge, the first ET evaluation of the quartet phase shifts {\it above} the breakup threshold. Overall, they they find good agreement with the results of full Faddeev calculations. Their most ambitious calculations include explicit pion loop contributions which precludes any comparison with the present work. They also show what they refer to as ``perturbative NLO calculations with pions integrated out". We have not established how to compare these results with the present work. However, they also show ``pionless NNLO" calculations -- based on a formulation appearing in Ref.~\cite{Bedaque:1998mb} -- which correspond to using our Eq.~\ref{f3ETFull} after setting $v_4'$ and all higher order terms to zero. Our attempts to reproduce these specific calculations reveal that they are subject to problems arising from spurious poles as discussed in connection with Eqs.~\ref{f3ETFull} and~\ref{fINVERE} in the preceding Section and as noted in Ref.~\cite{Bedaque:1998mb}, too. Our comparable $s$-wave results are very similar to those of Bedaque and Griesshammer~\cite{Bedaque:1999vb} and, in fact, are quite similar to our FF calculations appearing in Figs.~\ref{FR_Re_S} and~\ref{FR_Im_S} where we show the real and imaginary parts of the phase shifts, respectively. As these figures show, such calculations agree well with full Faddeev calculations. However, for this case we calculate small negative imaginary s-wave phase shifts below threshold (they are much larger than our numerical uncertainties) in violation of unitarity. Moreover, using the same approach to calculate $p$-wave phase shifts results in catastrophic failure in the form of large negative imaginary phase shifts at all energies. Both problems depend critically on the upper limit of the momentum integration in the relevant Faddeev equation.  Since the spurious pole referred to above appears at $k\simeq$1 fm$^{-1}$, the problem with negative imaginary phase shifts disappears if the upper limit of momentum is less than 1 fm$^{-1}$ so that the ``bad pole" cannot contribute. This dependence on the upper cutoff on momentum is undesirable and, in any case, is consistent with the discussion of the limitations of ET calculations as discussed above in connection with Eqs.~\ref{f3ETFull} and~\ref{fINVERE}.

The second of these papers which present ET-ERE calculations which do not rely on the prescription implied by Eq.~\ref{fINVERE} is the work of Griesshammer in Ref.~\cite{Griesshammer:2004pe} which makes use of the so-called ``$Z$-expansion" method, originally suggested in the 2-body context by Phillips, Rupak and Savage~\cite{Phillips:1999hh}. The essential feature of Griesshammer's method 
is to use (see Eq. 2.12 in Ref.~\cite{Griesshammer:2004pe}),
instead of the fully summed form of Eq.~\ref{f3ETFull},
\begin{equation}
{\rm f}(p)\rightarrow \frac{1}{-\gamma-ip}\cdot  \frac{1}{1-\frac{(Z_B-1)}{2\gamma}(\gamma+ip)},
\label{fZFull}
\end{equation}
the expanded form of Eq.~\ref{fINVERE}, 
\begin{equation}
{\rm f}(p)\rightarrow\frac{1}{-\gamma-i p}\biggl[1+\sum_{n=1}^{N_{max}}\biggl(\frac{(Z_B-1)}{2\gamma}\cdot (-\gamma-i p) \biggr)^n\ \biggr],
\label{fZERE}
\end{equation}
where $Z_B$ is the wave function renormalization appearing in Eqs.~\ref{ZUPA} and \ref{ZERENNLO}. 
The $Z$-expansion quartet $s$-wave imaginary phase shifts (Griesshammer does not show these phase shifts) at NLO and NNLO using Eq.~\ref{fZERE} oscillate wildly and are negative over much of the region from just above breakup to $p$= 200 MeV. Results using the fully summed form of Eq.~\ref{fZFull}  are satisfactory  and quite similar to our FF calculations appearing in Figs.~\ref{FR_Re_S} and~\ref{FR_Im_S} {\it provided}  the upper limit of momentum is less than 1 fm$^{-1}$ so that the spurious pole  in Eq.~\ref{fZFull} cannot contribute.  Otherwise, small negative imaginary phase shifts are again observed below threshold. Our calculations for the $Z$-expansion quartet $p$-wave phase shifts at LO, NLO and NNLO using Eq.~\ref{fZERE} agree well with those appearing in the Griesshammer paper and are consistent with our Figs.~\ref{FR_Re_P} and~\ref{FR_Im_P}.  However, the fully summed imaginary phase shifts are again large and negative at all energies unless the upper limit of momentum is less than 1 fm$^{-1}$ so as to eliminate the spurious pole contribution.

In closing this Section, we conclude that the ET-ERE approaches we have discussed in this paper, which ignore the finite range of the 2-body interaction at the outset and then try to accommodate it approximately via expansions based on Eq.~\ref{fINVERE}, face a difficult challenge. We conclude that, in this regard at least,  our FF formulation provides a superior starting point for the treatment of 3-body physics. 
We observe that finite range effects are relatively large for the case of quartet $nd$ scattering considered here since the expansion parameter $\gamma r_0\simeq 3\gamma/\beta\simeq 0.45$ is relatively large. In the doublet channel with both singlet and triplet contributions, the same physics occurs in the NN triplet contribution, but there is also a comparable contribution from the NN singlet channel where $\gamma r_0\rightarrow r_0/a_2\simeq -0.12$.  Even more advantageous is the situation for atomic helium where $r_0/a_2\simeq 0.075$. However, even in these cases, finite range effects can be important and hard to account for using, {\it e.g.}, the ET-ERE prescription when calculating observables such as 3-body recombination coefficients which necessarily involve momentum transfers large enough to be above the breakup  threshold and which depend on the imaginary part of the elastic phase shifts~\cite{shepard:062713}. 

It also important to observe that, in spite of the significant differences between them, both the full FF and the ET-ERE calculations use the same {\it on-shell} 2-body input, namely parameters fit to the deuteron binding energy and low energy phase shifts. However, the three-body kernel probes the 2-body amplitude off-shell.  Specifically, the form factor treatment provides momentum dependence in the ``blobs" of the ping-pong diagram (Fig.~\ref{ZDiag}) which softens the integral in a manner consistent with the 2-body phenomenology which results in a more accurate treatment of the off-shell behavior of the 2-body $T$-matrix than does the ET-ERE formulation without such regulating terms.  We have examined alternative parameterizations for the form factors, such as Gaussians~\cite{Kohler:2003,Stoll:2005,Kohler:2006zz}, and found that the specific analytic form of the form factors -- and the off-shell behavior they imply -- is not important provided the 2-body phenomenology is respected.

\section{Summary and Conclusions}
\label{sec:conclusions}
 
\begin{figure}[ht]
\vspace{0.50in}

\includegraphics[width=4in,angle=0]{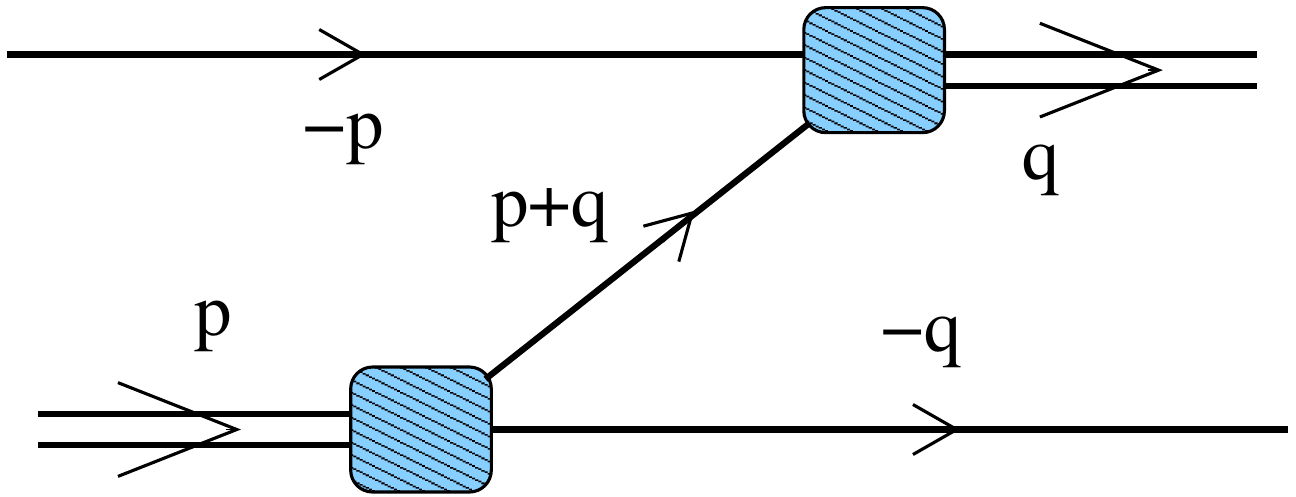}
\caption{(Color online).
         Diagrammatic representation of $\cal Z$, the Born (``ping-pong") term of 3-body particle-dimer elastic scattering Faddeev equation. The `` blobs'' are the form factors which depend on the magnitudes of the relative 2-body momenta which are $|{\bf p}/2+{\bf q}|$ and $|{\bf p}+{\bf q}/2|$ for the lower and upper form factors, respectively.}
\label{ZDiag}
\end{figure}

In this work we have studied the convergence properties of finite range expansions of 3-body scattering amplitudes by starting with a simple separable form factor model. The full (unexpanded) form factor calculations give an excellent account of 3-body $^4S_{3/2}$ $nd$ $s$- and $p$-wave phase shifts and other observables, too.  (For simplicity, we have ignored the $^2S_{1/2}$ channel in the present paper. Calculations to be presented in a future publication show that full FF doublet calculations similar to those presented here for quartet scattering are in similarly good agreement with the relevant full Faddeev calculations.) We developed a systematic expansion of the 3-body scattering amplitude in the inverse of the form factor parameter, $\beta$, which is  proportional to the range of the 2-body interaction. The resulting 3-body equations were referred to as FF-ERE expressions. We then sketched a derivation of the standard Effective Theory form of the 3-body Faddeev equations in which  range effects are included using the Effective Range Expansion for the 2-body scattering. We were then able to compare directly 
these two expansions at Leading Order (LO), Next-to-Leading Order (NLO) and Next-to-Next-to-Leading Order (NNLO).  Perhaps the most significant difference between the two approaches is that the ET-ERE formulation lacks the momentum dependence in the ``blobs" in the ping-pong diagram (Fig.~\ref{ZDiag}) arising from the form factors. 

Comparisons of numerical results  with full Faddeev calculations using realistic modern $NN$ interactions revealed, first of all, that our full Form Factor calculations  agree very well with the former results in spite of the simplicity of our dipole form factor. Both expansions, namely the FF-ERE and the ET-ERE, showed serious deficiencies, the most worrisome of which were large negative imaginary phase shifts at NLO and NNLO, a situation incompatible with unitarity.  This particular shortcoming was only overcome in the FF-ERE case by going to  N$^4$LO and even then, agreement with the full Faddeev results was poor. Poor convergence for both expansions was observed for $p$-wave phase shifts, too. We conclude that approaches which do not treat 2-body finite range effects directly and employ expansions such as those used in the ET-ERE or FF-ERE may be unreliable. Moreover, the technical problem frequently encountered in any attempt to include 2-body range effects {\it to all orders}, namely the appearance of spurious poles in ${\rm f}(p)$ -- as discussed in detail in Section 3.4 of Ref.~\cite{Griesshammer:2004pe} -- never appears in the full FF formulation as it is simple to show that there is only one pole corresponding to the single bound state found for any separable 2-body potential of a reasonable form.

The success of our simple full Form Factor results, especially in the regime where the standard ERE expansion breaks down, is gratifying and suggests that the approach warrants further study. We first note that, upon writing our dipole form factor as a geometric series
\begin{equation}
1/(1+q^2/\beta^2)=1-q^2/\beta^2-q^4/\beta^4-\dots,
\label{FFExp}
\end{equation}
we see that it automatically sums a certain set of contact terms to all orders! It is therefore perhaps not surprising that our simple FF calculations behave so differently from those based on the expansions discussed above. 

In the context of our Form Factor calculations, there is a 
fundamental question of whether the simple single form factor method (the Unitary Pole Approximation) we have employed here can be established as the first step in a sequence of controlled approximations. Preliminary work strongly suggests that we can answer this question in the affirmative. In a separate paper, we will use the Harms' expansion of an arbitrary potential via Eq.~\ref{vFFexp}, namely,
\begin{equation}
V=-\sum_{n=1}^N\ |\psi_n\rangle \lambda_n \langle\psi_n|,
\end{equation}
to construct a systematic method of improving upon the UPA by using separable potentials of successively higher rank, a procedure facilitated by the fact that the Harms expansion automatically orders terms by their importance at low momentum scales. Such a procedure is satisfactory from a fundamental point of view since the Harms expansion of the potential given above is exact in the limit that the sum includes all eigenkets of Eq.~\ref{twoFFeqn}. The practical utility of the method depends, of course,  on the convergence properties of this expansion. Again, preliminary work shows that, {\it e.g.}, for realistic potentials describing the $^1S_0$ $NN$ channel, only two or three terms of the Harms expansion are required for a very accurate reproduction of the on- and off-shell $T$-matrix for momenta less than $\sim$ 2 fm$^{-1}$. This convergence can be further enhanced by making Harms' expansions of $V_{low}$'s determined from realistic $NN$ potentials using, {\it e.g.}, Similarity Renormalization Group methods~\cite{Glazek:1993,Wegner:1994,Glazek:1994,Bogner:2006pc}. We will also have to extend our method to treat the tensor interactions in the full $^3S_1$-$^3D_1$ channel. Moreover, our 3-body calculations using these separable interactions must be extended to the $nd$ doublet channel and to the case of three identical spinless bosons ({\it e.g.}, $^4$He atom-dimer scattering) where Efimov physics is present. All of these points will be addressed in a future publication.


\smallskip
\acknowledgments
One of the authors (JAM) thanks the Physics Department of the University of Colorado, Boulder for hosting him during his sabbatical where this work was initiated.  This work was supported in part by DOE grant DE-FG02-93ER40774.

\bibliography{../References.bib}

\vfill\eject
\begin{table}
\begin{tabular}{|l|c|}
\hline
~~~~~~~Model & ~~~~~$^4a_3$ (fm)~~\\
 \hline
Experiment& 6.35$\pm$0.02\\
Full Faddeev& 6.325\\
LO& 5.091 (19.7)\\
ET-ERE-NLO& 5.935 (6.3)  \\
ET-ERE-NNLO& 6.201 (2.2) \\
FF-ERE-NLO& 6.089 (4.0)\\
FF-ERE-NNLO& 6.280 (0.9)\\
Full FF& 6.322 (0.05)\\
\hline
\end{tabular}
\caption{Calculations of $nd$ quartet scattering length, $^4a_3$, for various calculations are compared with an (old) experimental value~\cite{Dilg:1971} and results of full Faddeev calculations using modern NN interactions~\cite{Witala:2003mr}. Percent differences relative to the full Faddeev value appear in parentheses.}
\label{Table2}
\end{table}

\begin{table}
\begin{tabular}{|l|c|c|c|c|}
 \hline
~~~~~~~Model & ~~~~~$\sigma_{tot}$ (mb)~~& ~~~~~~$\sigma_{el}$ (mb)~~ &~~~~~ $\sigma_{BR}$ (mb)~~ &~ $\frac{d\sigma}{d\Omega}(\theta=180^o)$ (mb)\\
 \hline
ET-ERE-NNLO& 694.8 (85.5) & 673.9 (87.5)& 20.8 (49.4)& 152.3 (82.3) \\
FF-ERE-NNLO& 684.9 (84.3)& 652.4 (84.7)& 37.5 (89.1)& 160.8 (86.9)\\
Full Form Factor& 733.8 (90.3)& 695.2 (90.3)& 38.6 (91.7)& 172.1 (93.0)\\
Koike \&Taniguchi \cite{Koike:1986}& 812.3 & 770.2 & 42.1 & 185.0 \\
\hline
\end{tabular}
\caption{Calculations of center of mass total, elastic, break-up, and back-angle cross sections for quartet nd scattering at $E_{lab}=10$ MeV, summing partial waves up to $\ell=8$. Percentages of the Koike and Taniguchi~\cite{Koike:1986} values are in parentheses. }
\label{Table1}
\end{table}

\hfill\vfill\eject


\end{document}